\documentclass[twocolumn,showpacs,preprintnumbers,amsmath,amssymb]{revtex4}

\usepackage{graphicx}
\usepackage{dcolumn}
\usepackage{bm}
\usepackage{hhline}
\usepackage{color}

\begin{document}


\title{Ferromagnetic features on zero-bias conductance peaks in ferromagnet/insulator/superconductor junction}

\author{Nobukatsu Yoshida}%
 \email{yoshida.nobukatsu@nihon-u.ac.jp}
 \affiliation{Department of Liberal Arts and Basic Sciences, College of Industrial Technology, 
Nihon University, 2-11-1 Shin-ei, Narashino, Chiba 275-8576, Japan}

\author{Masashi Yamashiro}
 \email{yamashiro.masashi@nihon-u.ac.jp}
 \affiliation{Department of Liberal Arts and Basic Sciences, College of Industrial Technology, 
Nihon University, 2-11-1 Shin-ei, Narashino, Chiba 275-8576, Japan}

\date{\today}

\begin{abstract}
We present a general formula for tunneling conductance in ballistic 
ferromagnet/ferromagnetic insulator/superconductor junctions where 
the superconducting state has opposite spin pairing symmetry. 
The formula can involve correctly a ferromagnetism has been induced 
by effective mass difference between up- and down-spin electrons. 
Then, this effective mass mismatch ferromagnet and 
standard Stoner ferromagnet have been employed in this paper. 
As an application of the formulation, we have studied 
the tunneling effect for junctions including spin-triplet $p$-wave superconductor, 
where we choose a normal insulator for the insulating region whereas our 
formula can treat a ferromagnetic insulator. Then, we have been able to devote 
our attention to features of ferromagnetic metal.
The conductace spectra show a clear difference between two ferromagnets 
depending upon the way of normalization of the conductance.
Especially, a essential difference is seen in zero-bias conductance 
peaks reflecting characteristics of each ferromagnets. 
From obtained results, it will be suggested 
that the measurements of the tunneling conductance 
in the junction provide us a useful information 
about the mechanism of itinerant ferromagnetism in metals. 
\end{abstract}

\maketitle

\section{\label{sec1}
Introduction}

Andreev reflection(AR), which occurs at the interface 
of the junctions involving superconductors, is one 
of the most important elemental processes in 
the transport through the superconducting junctions \cite{ar}. 
A theory of transport taking into account the AR was formulated 
by Blonder, Tinkham and Klapwijk referred to as BTK theory\cite{btk}. 
The BTK theory enables us to probe the pairing state of superconductors. 
For example, for a junction consists of a normal metal and 
an unconventional superconductor, the quantum interference 
effect between the injected and the Andreev reflected particles, which feel mutually 
the different sign of superconducting pair potential 
through the scattering event at the interface of the junction, 
forms the so-called zero-energy Andreev bound states(ZABS)\cite{hu,yt1}. 
Indeed, the ZABS originated from the $d$-wave symmetry 
of superconducting pair potential have been observed as the zero-bias 
conductance peaks(ZBCPs) in tunneling experiment of 
high $T_{C}$ cuprate superconductors \cite{sk1,sk2,gd} 
according to theoretical prediction of Tanaka and 
Kashiwaya(TK) formula\cite{yt1}. 
The ZBCPs reflecting the ZABS are the 
essential feature of the electrical conductions
in normal metal/insulator/unconventional superconductor junction 
and provide us the information of the pairing symmetry of 
superconductor\cite{ym,hc,lof}. 
On the other hand, in ferromagnet/insulator/conventional 
superconductor junction, 
the measurements for low energy transport via AR offer the opportunity 
to probe the magnetic property such as the polarization of 
ferromagnetic materials\cite{jg,so,upa}. 
The AR in this junction is suppressed by the exchange field 
in the ferromagnet layer. 
As a result, the conductance at low energy of the junction is suppressed 
responding to the polarization of ferromagnet. 
The behavior of ZBCPs in ferromagent/insulator/unconventional 
superconductor junction 
have been studied to understand the characteristic properties of unconventional 
superconductors and to utilize the properties as applications of spintronics\cite{ev1,ev2,ev3}. 
In these junctions, 
the ferromagnet has been described within the Stoner model based 
on a picture of free electrons. 
However, in some materials, the other descriptions of 
ferromagnetism are required. 
The ferromagnetism kinetically driven 
by a spin-dependent bandwidth asymmetry, or, equivalently, 
by an effective mass splitting between $\uparrow$- and $\downarrow$-spin particles
\cite{sba1,sba2,sba3,sba4,sba5,sba6} is an interesting model 
giving itinerant ferromagnet. 
\par
Recently, Annunziata, et al. analyzed the charge and spin transport 
in ferromagnet/insulator/superconductor(F/I/S) 
junctions with taking account 
above mentioned the spin-dependent bandwidth asymmetry 
ferromagnet (SBAF) making the effective masses have different values in ferromagnet\cite{gae1,gae2,gae3}. 
They clarified that from the knowledge of 
the critical transmission angle the measurement 
of the effective mass difference among the particles 
could be possible\cite{gae1}. 
In there, it has been also shown that the F/I/S junction 
is an effective probe to investigate the mechanism of ferromagnetism  
and the pairing symmetry of unconventional superconductor. 
Furthermore, it is suggested that the F/I/S junction 
can be useful as switching device using the spin current 
in the case of the symmetry of superconductor is conventional $s$-wave case\cite{gae2} 
and be possible as a spin-filtering device\cite{gae3}. 
They have studied on F/I/S 
junctions with several types of superconducting symmetries as 
the conventional $s$-wave, unconventional $d_{x^2-y^2}$-wave,  
and the time reversal symmetry broken $d_{x^2-y^2}+is$ or 
$d_{x^2-y^2}+d_{xy}$ states. 
Moreover, although zero bias anomaly of differential resistance has been shown 
experimentally in Sr$_{2}$RuO$_{4}$-Pt point contact experiment\cite{lau}, 
more recently, Kashiwaya et. al.\cite{sk3} has shown the ZBCP which is direct evidence of the 
ZABS by tunneling spectroscopic experiment of Sr$_{2}$RuO$_{4}$-Au junction.
There is much interest and importance to 
investigate problems on spin triplet $p$-wave symmetry nature 
because the $p$-wave pairing, especially, 
a chiral $p_{x} \pm ip_{y}$ state breaking time-reversal symmetry 
is one of the best candidates for 
bulk superconducting state of Sr$_{2}$RuO$_{4}$
\cite{sk3,mae1,ish,luke,mae2,mae3,sig,mae4}.
\par
In this paper, a formulation of the tunneling conductance for charge 
and spin currents in ferromagnet/ferromagnetic-insulator/superconductor 
(F/FI/S) junctions will be presented by taking the effective 
mass difference leading the spin-band 
asymmetry between $\uparrow$- and $\downarrow$-spin 
particles in ferromagnet \cite{gae1,gae2,gae3} 
into our previous theory\cite{ev1}. 
Although the formulation can be used for singlet and $S_{z}=0$ 
triplet superconductors, we will study a chiral $p$-wave state. 
Our formula has general form being able to 
include the ferromagnetic insulator, however, a normal insulator 
surrounded by the ferromagnet and the superconductor is considered 
for the insulating layer to get pure characteristic features of a ferromagnetism 
in a ferromagnet.
It is found that the normalized conductance spectra shows a clear 
difference between Stoner and spin-band asymmetry ferromagnets 
(STF and SBAF for abbreviation). 
The present results may be helpful 
in investigations of the mechanism of ferromagnet. 
\par
This paper is organized as follows. 
In Sec.I\hspace{-.1em}I we explain a theoretical model and 
derive a formulation following our previous method based on the BTK theory. 
The results for ferromagnet/insulator/chiral $p$-wave superconductor 
junctions are presented in Sec.I\hspace{-.1em}I\hspace{-.1em}I. 
Finally the results are summarized in Sec.I\hspace{-.1em}V.

\section{\label{sec2}
Model And Formulation}

For the model of formulation, we consider 
a two-dimensional ballistic F/FI/S junction with semi-infinite electrodes 
shown in figure 1. A flat interface is assumed to be located 
at $x$=0, and the ferromagnetic insulator for up(down) spin 
is described by a potential 
$V_{\uparrow(\downarrow)}(x)$ = $( V_{0} +(-) V_{ex} )\delta(x)$, 
where $\delta(x)$, $V_{0}$ and $V_{ex}$ are the $\delta$ function, 
a nonmagnetic barrier amplitude and a magnetic barrier amplitude, 
respectively. 
\begin{figure}
\begin{center}
\includegraphics[width=6cm]{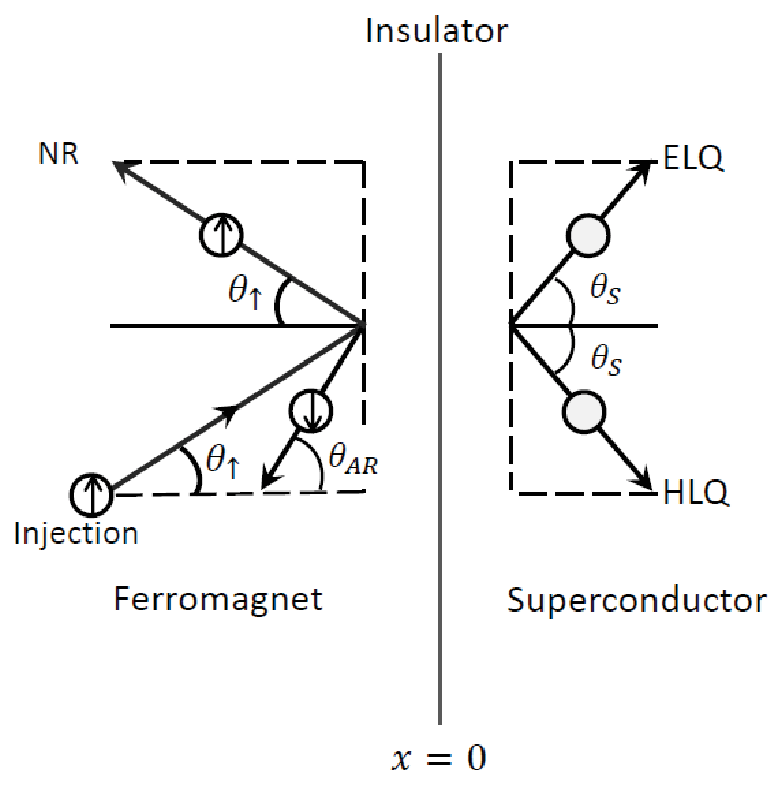}
\end{center}
\caption{Schematic illustration of scattering processes of an injected electron with $\uparrow$-spin 
at the F/FI/S ballistic junction.  Here, $\theta_{\uparrow}$, $\theta_{AR}$, and $\theta_{S}$ are 
injection, Andreev reflection, and transmission angles, respectively. It is assumed that the normal reflection 
at the interface is totally specular, then normal reflection angle is also given by $\theta_{\uparrow}$. 
For the case of $\downarrow$-spin electron, it can be depicted by 
flipping $\uparrow$ by $\downarrow$ in the figure.}
\label{fig1}
\end{figure}

\begin{figure}
\begin{center}
\includegraphics[width=7cm]{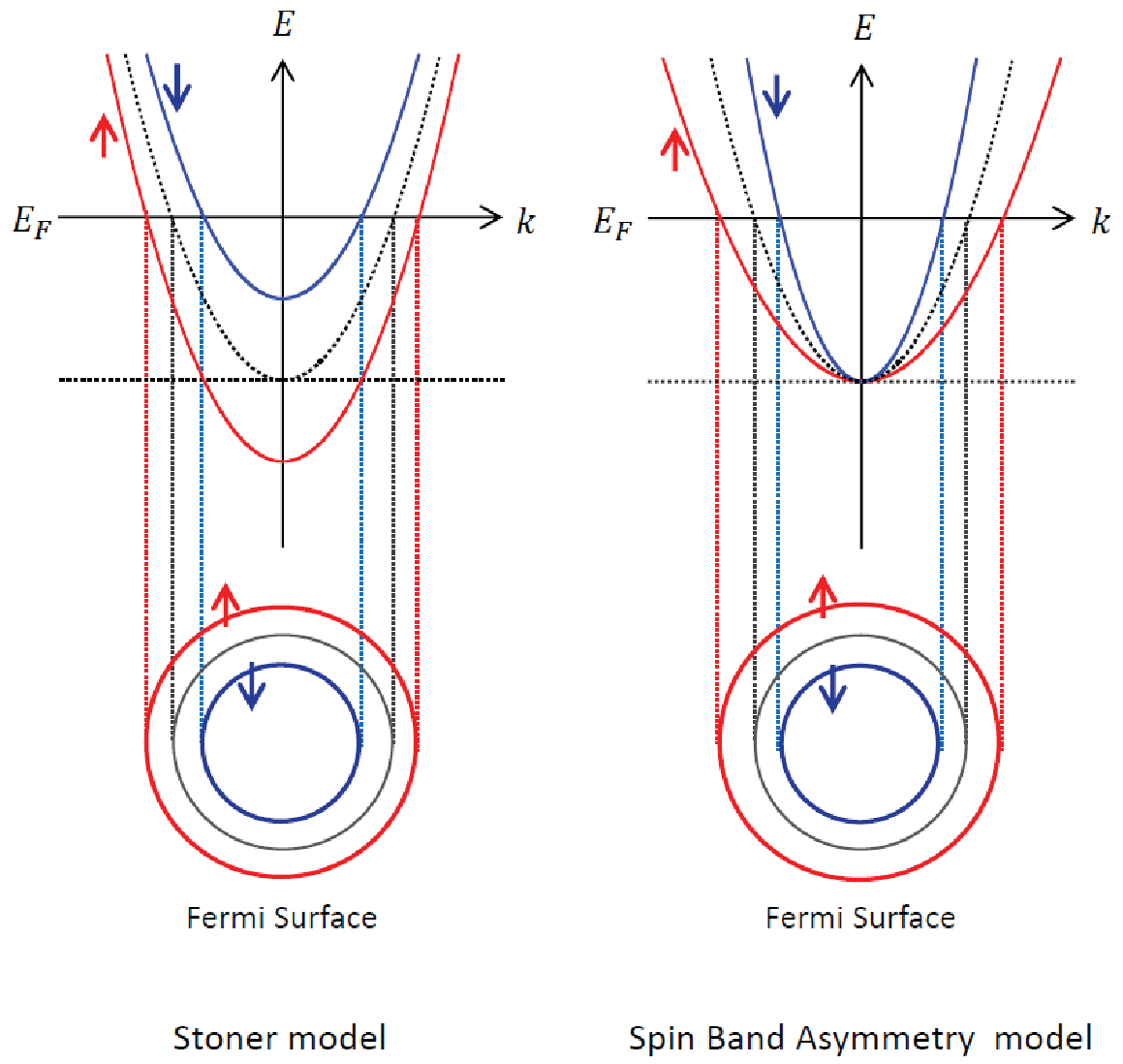}
\end{center}
\caption{A sketch of dispersion relation between energy and wave number, and 
Fermi surface for STF (left side) and for SBAF (right side). It is assumed that 
free electron model with rigidly energy shift in STF and different effective masses for each spin in SBAF.}
\label{fig2}
\end{figure}

For the ferromagnetism in the F electrode, we adopt two kinds models of
mechanisms shown in Fig\ref{fig2}.
One of these is the standard Stoner model in which the ferromagnetism is
induced
by the exchange potential leading to the rigid energy shift between
$\uparrow$-spin and
$\downarrow$-spin bands. The other is a spin bandwidth asymmetry model proposed by
Hirsch\cite{sba2},
in which the bandwidth is tuned relatively by the ratio of the effective
masses between
$\uparrow$- and $\downarrow$-spin particles. 
Although in the following the free particle-like
spectra of parabolic type is assumed as normal electronic dispersion
relation, we suppose to define 
the concept of bandwidth for the above description.
It implies that there can be some relations between this description and
some effective one-band tight binding model permitting 
the effective masses of carriers being
proportional to the inverse of the width of the bands where the carriers
get itinerancy.
Hence, only giving different values of the masses for $\uparrow$- 
and $\downarrow$-spin electrons yields a bandwidth asymmetry.

The spatial dependence of the pair potential is taken as 
$\Delta(\mbox{\boldmath $r$}) = \Delta\Theta(x)$ for simplicity. 
In addition, we consider the $S_z = 0$ pairing states, where 
the elements of pairpotential are given by 
$\Delta_{\uparrow,\uparrow}=\Delta_{\downarrow,\downarrow}=0$ 
and $\Delta_{\uparrow,\downarrow} = - \Delta_{\downarrow,\uparrow}$ 
for the singlet pairing state or 
$\Delta_{\uparrow,\downarrow} = \Delta_{\downarrow,\uparrow}$ 
for triplet pairing state.
Thus, the effective Hamiltonian(Bogoliubov-de Gnnes (BdG) equation) of the system 
can be reduced the decoupled equation 
for the eigen states 
$(u_{F(S)\uparrow(\downarrow)}(\mbox{\boldmath $r$}),
v_{F(S)\downarrow(\uparrow)}(\mbox{\boldmath $r$}))^T$
and is given by 

\begin{eqnarray}
\left(
\begin{array}{cc}
H_0^\sigma(\mbox{\boldmath $r$}) & \Delta(\mbox{\boldmath $r$}) \\
 \Delta^*(\mbox{\boldmath $r$}) 
& -H_0^{\bar{\sigma}}(\mbox{\boldmath $r$}) \\
\end{array} \right)
\left( \begin{array}{cc} u_{F(S)\sigma}(\mbox{\boldmath$r$}) \\ 
v_{F(S)\bar{\sigma}}(\mbox{\boldmath$r$}) \\ \end{array} \right)
= E\left( \begin{array}{cc} u_{F(S)\sigma}(\mbox{\boldmath$r$}) \\ 
v_{F(S)\bar{\sigma}}(\mbox{\boldmath$r$}) \\ \end{array} \right).
\end{eqnarray}
Here $E$ is the energy of the quasipaticle 
and $H_{0}^{\sigma}(\mbox{\boldmath $r$})$ is 
the single particle Hamiltonian for $\sigma$-spin 
where $\bar{\sigma}= -\sigma$. 
In the Ferromagnet side, the single particle Hamiltonian 
is given by $H_0^\sigma(\mbox{\boldmath $r$}) = 
-\hbar^2\nabla^2/2m_\sigma - \rho U_{ex} - E_{FM}$ 
where $\sigma= \uparrow,\downarrow$, $m_\sigma$ is the effective 
mass for $\sigma$-band particle, 
$\rho= +1(-1)$ for $\uparrow(\downarrow)$-spin, 
$U_{ex}$ is the exchange potential and $E_{FM}$ is the Fermi energy. 
The $H_0^\sigma$ in the superconductor side is 
given by $H_0(\mbox{\boldmath $r$}) = H_0^\sigma(\mbox{\boldmath $r$}) = 
-\hbar^2\nabla^2/2m_S- E_{FS}$ where the $m_s$ and $E_{FS}$ 
are the effective mass of the quasiparticle and the Fermi energy, respectively.
To describe the Fermi surface difference, we assume $E_{FM} = E_{FS} = E_{F}$. 

In the following, we apply the quasiclassical approximation where 
$E_F$ $\gg$ ( E, $\Delta(\mbox{\boldmath $k$})$ ) and 
the $k$-dependence of $\Delta(\mbox{\boldmath $k$})$ is replaced by 
the angle $\theta_S$ between the direction of the trajectory of quasiparticles 
in the superconductor and the interface normal. 
In the quasiclassical approximation, 
the wave vectors of $k_{\uparrow(\downarrow)}$ and $k_{ELQ(HLQ)}$ are given by 
$k_{\uparrow(\downarrow)}=
\sqrt{(2m_{\uparrow(\downarrow)}/\hbar^{2})(E_{F}+(-)U_{ex})}$ and 
$k_{ELQ(HLQ)}=k_{S}=\sqrt{2m_{S}E_{F}/\hbar^{2}}$, respectively, where 
ELQ (HLQ) indicates electronlike (holelike) quasiparticles. 
For example, we assume the injection of $\uparrow$-spin electrons 
from the ferromagnet at an angle $\theta_{\uparrow}$ to the 
interface normal as shown in Fig.\ref{fig1}. 
There are four possible scattering trajectories exist; 
Andreev reflection with angle $\theta_{AR}$ as holes belonging 
to $\downarrow$-spin band, 
normal reflection(NR), transmission to superconductor 
as ELQ, and transmission as HLQ. 
These four processes are described in same way for $\downarrow$-spin electrons 
with changing the scattering angle $\theta_{\uparrow}$ to $\theta_{\downarrow}$. 
Since the translational symmetry holds along the $y-$axis, 
the parallel momentum components of all trajectories are conserved 
$k_{\sigma}\sin\theta_{\sigma}$
=$k_{\bar{\sigma}}\sin\theta_{\bar{\sigma}}$
=$k_{S}\sin\theta_S$. 
The angles $\theta_{\sigma}$, $\theta_{\bar{\sigma}}$ and 
$\theta_{S}$ differ from each other except when
$U_{ex}=0$ and 
$m_{\uparrow}$=$m_{\downarrow}$=$m_S$, 
which means retroreflectiverly of AR broken by the exchange 
field and the effective masses difference. 
The BdG equations are reduced to 
the effective one-dimentional equation due to 
the translational invariance along $y$-axis of the Hamiltonian. 
Thus, the solutions of the BdG equations for $\sigma$-spin electron 
injections are described as  

\begin{widetext}
\begin{align}
\left(
\begin{array}{c}
u_{F\sigma}(x<0) \\
v_{F\bar{\sigma}}(x<0) \\
\end{array}
\right)
&=\left(
\begin{array}{c}
1 \\
0 \\
\end{array}
\right)e^{ik_{\sigma}x\cos\theta_{\sigma}}+
a_{\bar{\sigma}}\left(
\begin{array}{c}
0 \\
1 \\
\end{array}
\right)e^{ik_{\bar{\sigma}}x\cos\theta_{\bar{\sigma}}}+
b_{\sigma}\left(
\begin{array}{c}
1 \\
0 \\
\end{array}
\right)e^{-ik_{\sigma}x\cos\theta_{\sigma}}
\\[12pt]
\left(
\begin{array}{c}
u_{S\sigma}(x>0) \\
v_{S\bar{\sigma}}(x>0) \\
\end{array}
\right)
&=c_{\sigma}\left(
\begin{array}{c}
u_{+} \\
v_{+}e^{-i\phi_{+}} \\
\end{array}
\right)e^{ik_{S}x\cos\theta_{S}}+
d_{\sigma}\left(
\begin{array}{c}
v_{-}e^{i\phi_{-}} \\
u_{-} \\
\end{array}
\right)e^{-ik_{S}x\cos\theta_{S}}
\\
\intertext{with} 
u_{\pm}&=\sqrt{\frac{1}{2}\left( 1 + \frac{\Omega_{\pm}}{E}  \right)}, 
\hspace{12pt}
v_{\pm}=\sqrt{\frac{1}{2}\left( 1 - \frac{\Omega_{\pm}}{E}  \right)} 
\hspace{12pt}
\Omega_{\pm}=\sqrt{E^2 - \left|\Delta_{\pm}\right|^2}, \\[10pt]
e^{i\phi_{\pm}}&=\frac{\Delta_{\pm}}{\left|\Delta_{\pm}\right|}, 
\hspace{12pt}
\Delta_{+}=\Delta(\theta_{S}), \hspace{12pt} \Delta_{-}=\Delta(\pi-\theta_{S})
\end{align}
where the probability coefficients $a_{\bar{\sigma}}$, 
$b_{\sigma}$, $c_{\sigma}$, and $d_{\sigma}$ are for AR, NR, transmission ELQ and HLQ.
These coefficients are calculated from the boundary conditions at $x=0$,
\begin{gather}
u(v)_{F\sigma(\bar{\sigma})}(x=0)=u(v)_{S\sigma(\bar{\sigma})}(x=0), \\[12pt]
\displaystyle
\left.\frac{\hbar^{2}}{2m_{S}}
\frac{du_{S\sigma}}{dx}\right|_{x=0}-
\left. \frac{\hbar^{2}}{2m_{\sigma}}
\frac{du_{F\sigma}}{dx}\right|_{x=0}=
V_{\sigma}u_{S\sigma}(x=0) \\[12pt]
\left.\frac{\hbar^{2}}{2m_{S}}
\frac{dv_{S\bar{\sigma}}}{dx}\right|_{x=0}-
\left. \frac{\hbar^{2}}{2m_{\bar{\sigma}}}
\frac{dv_{F\bar{\sigma}}}{dx}\right|_{x=0}=
V_{\bar{\sigma}}v_{S\sigma(\bar{\sigma})}(x=0)
\end{gather}
\end{widetext}

As explained in our previous paper, the reflection process 
depends upon the size relation of the Fermi surfaces 
between FM and SC. 
In the following, we will consider a situation where 
$k_{\downarrow}<k_{S}<k_{\uparrow}$, and $m_{\uparrow}/m_{S}=m_{S}/m_{\downarrow}$ 
with $m_{\uparrow}>m_{S}>m_{\downarrow}$.
Following the BTK theory with taking care of the probability conservation 
of quasiparticle flow,  
\begin{equation}
\mid b_{\sigma} \mid^{2} + 
\frac{v_{f,\bar{\sigma}}}{v_{f,\sigma}} \mid a_{\bar{\sigma}} \mid^{2} 
+ \frac{v_{s,+}}{v_{f,\sigma}} \mid c_{\sigma} \mid^{2}
+ \frac{v_{s,-}}{v_{f,\sigma}} \mid d_{\sigma} \mid^{2}
= 1 
\end{equation}
the conductance $G_{S,\sigma}^{C(S)}$ for 
the $\sigma$-spin charge(spin) current through 
the system can be calculated by 
\begin{equation}
G_{S,\sigma}^{C(S)}
= 1 +(-) 
\frac{v_{f,\bar{\sigma}}}{v_{f,\sigma}} \mid a_{\bar{\sigma}} \mid^{2} 
- \mid b_{\sigma} \mid^{2} 
\end{equation}
where $v_{f,\sigma}=\hbar k_{\sigma}/m_{\sigma}$ is the group velocities of 
the $\sigma$-spin particles in ferromagnet and $v_{s,+(-)}=\hbar k_{S}/m_{S} $ 
is that of the ELQ (HLQ) in superconductor. 
It is much worth to note that our conductance formula
$G_{S,\sigma}^{C(S)}$ is different from former works\cite{gae1,gae2,gae3}. 
On our way of formulating the conductance as an extension of previous formulation
\cite{ym,ev1} to the present situation, it has to be needed for correct 
treatment of mass mismatch in a same metal that the cofficient of AR $a_{\bar{\sigma}}$ 
should be given by the ratio of group velocities 
rather than that of wavenumbers as a consequence of conservation 
law of particle flow. 
Using the obtained AR ($a_{\bar{\sigma}}$) and NR ($b_{\sigma}$) 
coefficients in the same way as the 
previous paper\cite{ev1} based on the TK formula\cite{yt1}, 
the charge(spin) conductance for each spin 
$\uparrow$ and $\downarrow$ can be formulated by  
\begin{widetext}
\begin{align}
G_{S,\uparrow}^{C(S)}&=
G_{N,\uparrow}
\frac{1-\mid \Gamma_{+} \Gamma_{-} \mid^{2}(1-G_{N,\downarrow})
+(-) G_{N,\downarrow} \mid{\Gamma_{+}} \mid^{2} }
{\mid 1 - \Gamma_{+} \Gamma_{-}
\sqrt{1-G_{N,\downarrow}} \sqrt{1-G_{N,\uparrow}} 
\exp[i(\varphi_{\downarrow}-\varphi_{\uparrow})] \mid^{2}} 
\Theta(\mid \theta_{S} \mid -\theta_{C})                      
\notag \\[12pt]
&
+[1-\Theta(\mid \theta_{S} \mid -\theta_{C})]
G_{N,\uparrow}
\frac{1-\mid \Gamma_{+} \Gamma_{-} \mid^{2}}
{\mid 1 - \Gamma_{+} \Gamma_{-}
\sqrt{1-G_{N,\uparrow}} 
\exp[i(\varphi_{\downarrow}-\varphi_{\uparrow})] \mid^{2}} 
\\[12pt]
G_{S,\downarrow}^{C(S)}
&=
G_{N,\downarrow}
\frac{1-\mid \Gamma_{+} \Gamma_{-} \mid^{2}(1-G_{N,\uparrow})
+ G_{N,\uparrow} \mid{\Gamma_{+}} \mid^{2}}
{\mid 1 - \Gamma_{+} \Gamma_{-}
\sqrt{1-G_{N,\downarrow}} \sqrt{1-G_{N,\uparrow}} 
\exp[i(\varphi_{\uparrow}-\varphi_{\downarrow})] \mid^{2}} 
\Theta(\mid \theta_{S} \mid -\theta_{C})                       
\\
\intertext{with} 
G_{N,\uparrow(\downarrow)}&=
\frac{4\lambda_{\uparrow(\downarrow)}}
{(1+\lambda_{\uparrow(\downarrow)})^{2}+Z_{\uparrow(\downarrow)}^{2}},
 \hspace{20pt} 
\exp(i\varphi_{\uparrow(\downarrow)})=
\frac{1 - \lambda_{\uparrow(\downarrow)} + iZ_{\uparrow(\downarrow)}}
{1 + \lambda_{\uparrow(\downarrow)} + iZ_{\uparrow(\downarrow)}}, 
\notag \\[12pt]
Z_{\uparrow(\downarrow)}&=\frac{Z_{0,\uparrow(\downarrow)}}{\cos\theta_{S}}, 
\hspace{20pt}  
Z_{0,\uparrow(\downarrow)}=\frac{2m_{s}(V_{0}-(+)V_{ex})}{\hbar^2k_{S}}, 
\hspace{20pt}  
\Gamma_{\pm}
= \frac{v_{\pm}}{u_{\pm}}, 
\notag \\[12pt]
\lambda_{\uparrow(\downarrow)}&=
\sqrt{ \gamma^{-1(1)} + \frac{\gamma^{-1/2(1/2)}}{\cos^{2}\theta_{S}} 
\left( 1-\gamma^{-1/2(1/2)} +(-) \chi  \right) }, 
\end{align}
\end{widetext}
where $\theta_{C}\equiv \cos^{-1}\sqrt{\gamma^{-1/2}(\chi-1+\gamma^{1/2})}$
( or $\sin^{-1}\sqrt{\gamma^{-1/2}(1-\chi) }$ ) is the critical 
angle of the AR measured in the superconductor side. 
Here, $\chi=U_{ex}/E_{F}(0\leq\chi\leq1)$ and $\gamma=m_{\uparrow}/m_{\downarrow}\geq1$. 
In the above, $G_{N,\sigma}$ corresponds to the conductance when 
the superconductor is in the normal state. 
We calculate the normalized conductance defined by 
\begin{equation}
G^{C(S)}_{T}(eV)=
\frac{ \int^{\pi/2}_{\pi/2} d\theta_{S} \cos\theta_{S}
(P_{\uparrow} G_{S,\uparrow}^{C(S)} + P_{\downarrow} G_{S,\downarrow}^{C(S)})}
{\int^{\pi/2}_{\pi/2} d\theta_{S}\cos\theta_{S}
(P_{\uparrow} G_{N,\uparrow} + P_{\downarrow} G_{N,\downarrow})}
\label{eq11}
\end{equation}
where the polarization $P_{\sigma}$ for $\sigma$-spin 
is expressed as 
\begin{equation}
P_{\uparrow}=\displaystyle
\frac{\gamma(1+\chi)}{\gamma(1+\chi)+1-\chi}, \hspace{18pt}
P_{\downarrow}=\displaystyle
\frac{1-\chi}{\gamma(1+\chi)+1-\chi}.
\nonumber
\end{equation}

It is noted in general that the normalized conductance will be defined 
alternatively corresponding to the actual experiments.

Above formulas (2.11), (2.12), and (2.14) can reproduce 
former formulas of tunneling conductance for junctions including triplet superconductor (TS). 
For $m_{\uparrow}=m_{\downarrow}$, 
these eqations coincide to that of STF/I/TS junction\cite{ev1}, and for 
$m_{\uparrow}=m_{\downarrow}$ and $U_{ex}=0$, the conductance formula for 
N/I/TS junction\cite{ym} is reproduced.

\section{\label{sec3}
Results}

At first, we notice about the growth of the magnetization $M$ for STF or SBAF. 
Using the polarization $P_{\sigma}$, $M$ is given by 
$M = P_{\uparrow}-P_{\downarrow}$. 
For pure STF case ($\gamma=1$), the magnetization is equal 
to the magnitude of exchange splitting 
$M= \chi$ $(=U_{ex}/E_{F})$. 
For pure SBAF case($\chi=0$), the $M$ is given 
by $M = (\gamma-1)/(\gamma+1)$.
 Thus, the half metal state in SBAF case is unphysical 
situation because $\gamma=\infty$. Figure 3 shows 
the $M$ in SBAF case as a function of $\gamma$. 
It can be seen that the growth rate of $M$ becomes very 
gradual over $\gamma \approx 50$. 
From this, one can expect the clear differences of transport properties depending on $M$ 
between STF and SBA near the half metallic limit. 
Hereafter, we call ``strong ferromagnetic regime'' as a region under and near 
the half metallic limit.

In the following sub sections, we apply our conductance formula to 
ferromagnet/insulator/triplet superconductor (F/I/TS) junction
(F referred to as STF or SBAF) where $V_{ex}=0$. 
As the pairing potential, a triplet $p$-wave state is employed 
by choosing $\Delta_{\uparrow\downarrow}(\theta_{S})=$
$\Delta_{\downarrow\uparrow}(\theta_{S})=$
$\Delta_{0}\exp(i\theta_{S})$, 
$\Delta_{\uparrow\uparrow}(\theta_{S})=$
$\Delta_{\downarrow\downarrow}(\theta_{S})=0$ 
for opposite spin pairing.
And in addition, we choose some sets of parameters $(\chi,\gamma)$ giving the same 
$M=\{0,0.25,0.5,0.75,0.99\}$ 
shown in Table \ref{tab1} so as to get clear characteristics of each ferromagnets.
\begin{table}
\caption{Numerical values of the magnetization $M$, the normalized exchange 
interaction $\chi=U_{ex}/E_{F}$, and the mass ratio 
$\gamma=m_{\uparrow}/m_{\downarrow}$. 
For the value $M=0.25$, there are two cases, one is pure STF, $\chi=0.25$ and $\gamma=1$, 
and the other is pure SBAF, $\chi=0$ and $\gamma=5/3$. 
All other values of $M$ are in the same way except 
the case of $M=0$.}
\begin{tabular}{ccc}\toprule
magnetization $M$ \hspace{15pt} 
& exchange Int.$\chi$ \hspace{15pt} & mass mismatch $\gamma$  \\ 
\colrule
0 \hspace{15pt} & 0 \hspace{15pt} & 1 \\ 
0.25 \hspace{15pt} & 0.25 \hspace{15pt} & 1  \\ 
0.25 \hspace{15pt} & 0 \hspace{15pt} & 5/3  \\ 
0.5 \hspace{15pt} & 0.5 \hspace{15pt} & 1  \\ 
0.5 \hspace{15pt} & 0 \hspace{15pt} & 3  \\ 
0.75 \hspace{15pt} & 0.75 \hspace{15pt} & 1  \\ 
0.75 \hspace{15pt} & 0 \hspace{15pt} & 7  \\ 
0.99 \hspace{15pt} & 0.99 \hspace{15pt} & 1  \\
0.99 \hspace{15pt} & 0 \hspace{15pt} & 200  \\
\botrule
\end{tabular}
\label{tab1}
\end{table}

\begin{figure}
\begin{center}
\includegraphics[width=5cm]{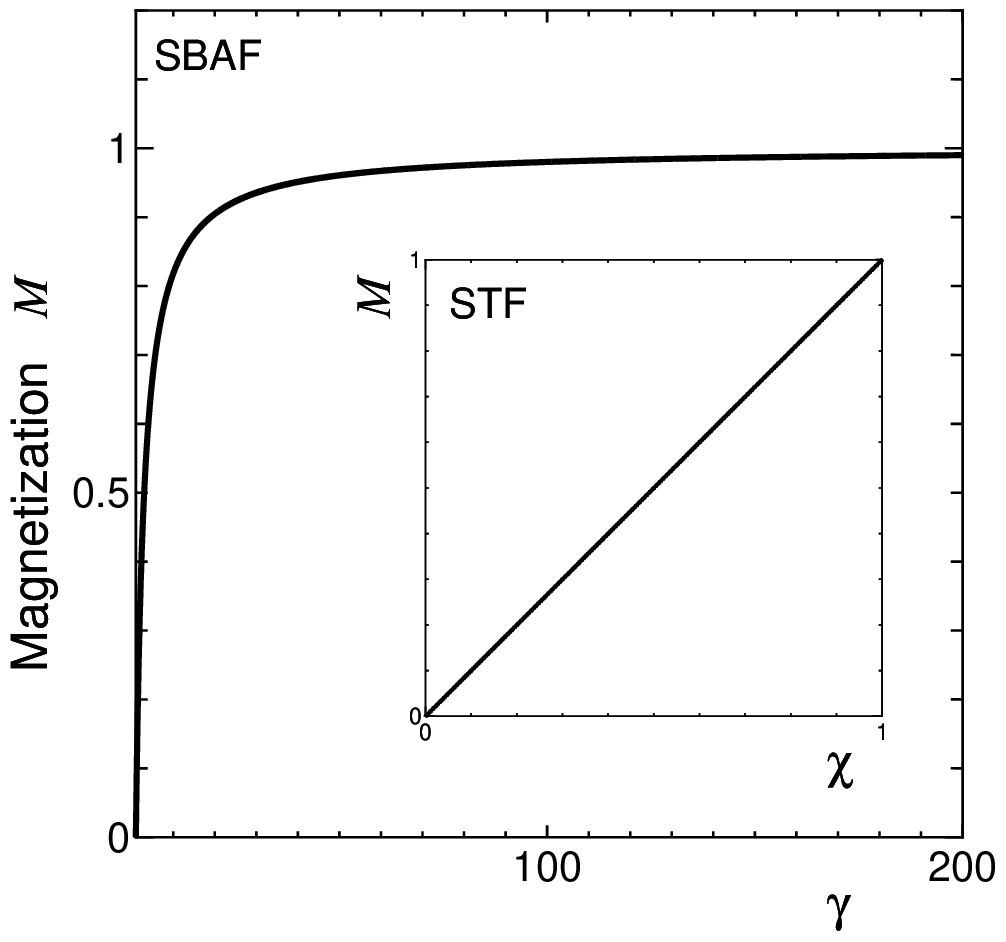}
\end{center}
\caption{The magnetization $M$ as a function 
of $\gamma$ in the pure SBAF, 
and inserted panel is $M$ as a function of $\chi$ in 
the pure STF.}
\label{fig3}
\end{figure}

\subsection{\label{sec31}
Distinction between STF and SBAF}

\begin{figure}
\begin{center}
\includegraphics[width=8.5cm]{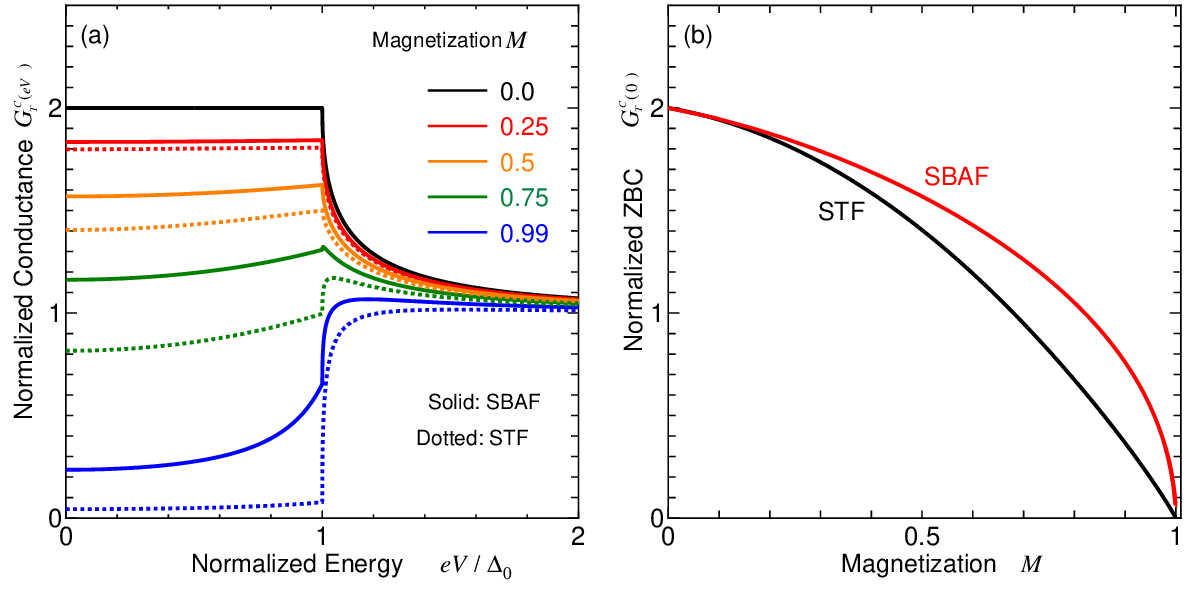}
\end{center}
\caption{ Normalized conductance spectra for the charge current $G_{T}^{C}(eV)$ in the metallic limit, $Z=0$ in (a), 
and in (b), height of conductance at zero energy is plotted 
as a function of the magnetization $M$ for 
STF (red line) and for SBAF (black line).}
\label{fig4}
\end{figure}

\begin{figure}
\begin{center}
\includegraphics[width=5cm]{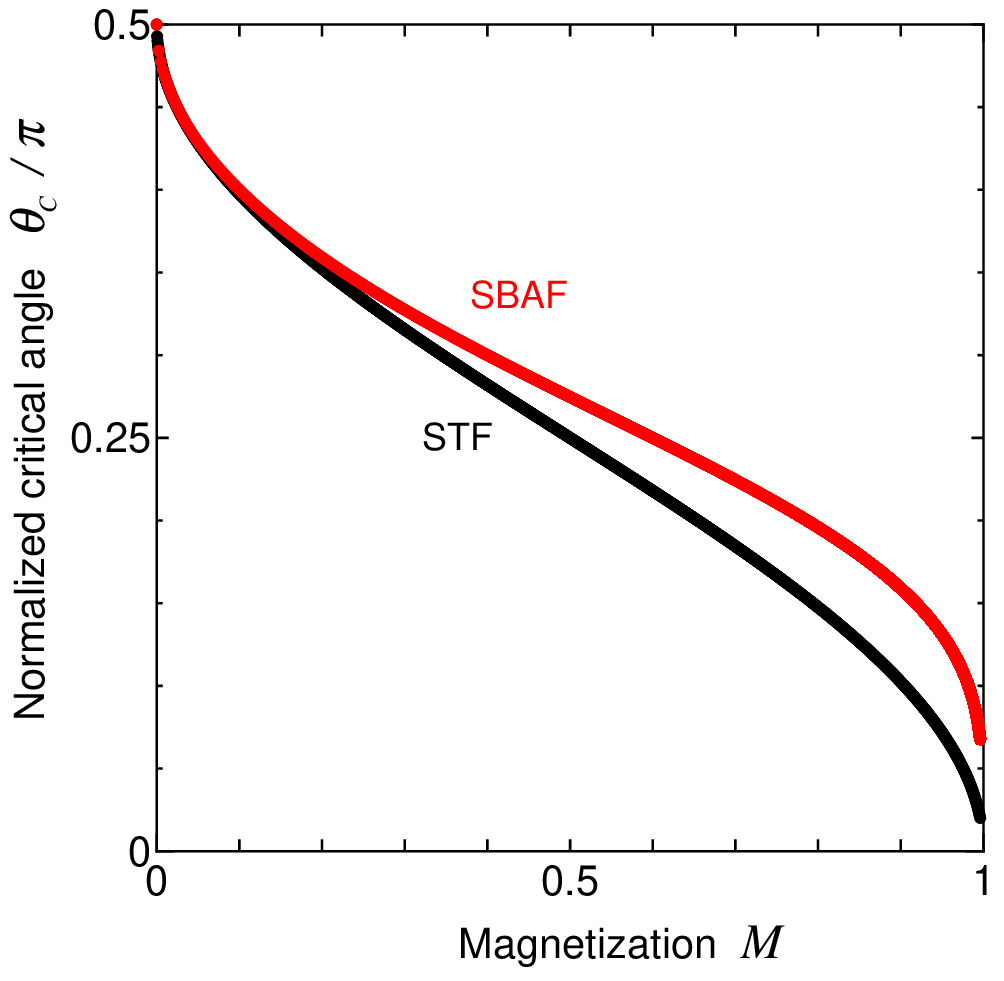}
\end{center}
\caption{ Critical angles of AR as functions 
of $M$ for STF (black line) and for SBAF (red line). 
As described in main text, these angles are determined by the conservation 
condition for momenta parallel to the interface, 
$k_{\downarrow}\sin\theta_{\downarrow}=k_{S}\sin\theta_{S}$. Such as in the present model, i.e., the ferromagnetism is given 
by a mismatch in kinetic energy the AR critical angle is determined by $\theta_{C}=\theta_{S}$ for the case of $\theta_{\downarrow}=\pi/2$ 
because of satisfying the condition $k_{\downarrow}<k_{S}$. }
\label{fig5}
\end{figure}

To investigate a consequence of the different mechanism 
of the magnetization, avoiding any effects of the normal barrier 
we consider the highly transparent junction in the metallic limit ($Z_{0}=0$). 
In this case, the normalized total conductances 
$G_{T}^{C}(eV)$ show same trend that conductance values 
inside the energy gap $eV < \Delta_{0}$ are reduced 
when the value of magnetization $M$ is increased for 
both STF and SBAF (Fig.\ref{fig4}(a)). 
It indicates that the retro-reflectivity of the AR is broken due 
to the induced $M$.
However, the $M$ dependence of reduction for 
$G_{T}^{C}(eV<\Delta_{0})$ is different for each of them. 
The difference can be seen more clearly in the $M$-dependence of 
conductance values at $eV=0$, $G_{T}^{C}(0)$, in Fig\ref{fig4}.(b).
It is found that the suppression of $G_{T}^{C}(0)$ for SBAF case is weaker 
rather than that for STF case without weak magnetization regime, $ 0.0 \leq M \simeq 0.2$ 
and at half metallic limit, $M=1.0$. 
To clear the reason of different $M$ dependence of conductances 
for SBAF case and STF case, we show the critical angle of AR as a function of $M$ in Fig.\ref{fig5}.  
The $\theta_{C}$ for both SBAF and STF cases decreases with increasing $M$. 
It is found that the difference between angles is getting larger from $M \sim 0.2$ to $\sim 0.9$, 
and converges to zero at $M=1.0$.
For nearly half metallic limit $M=0.99$, 
the $\theta_{C}$ is almost suppressed in the STF case, 
while there still remains in the SBAF case. 
The critical angles for STF and SBAF are  
$\theta_{C}=\cos^{-1}\sqrt{M}(=\sin^{-1}\sqrt{1-M})$ 
and $\theta_{C}=\cos^{-1}\sqrt{1-(\frac{1-M}{1+M})^{1/2}}
(=\sin^{-1}\sqrt{(\frac{1-M}{1+M})^{1/2}})$, respectively. 
Then, it is clear that $\theta_{C}$ in SBAF case is 
larger than that in STF case for same $M$ except non-magnetic state, 
$M=0$ and half metal state, $M=1$. 
Consequently, as shown in Fig.\ref{fig4}, the $G_{T}^{C}(eV<\Delta)$ 
in SBAF case is larger than that in STF case. 

\subsection{\label{sec32}Ferromagnetic feature on ZBCP}

\begin{figure}
\begin{center}
\includegraphics[width=5cm]{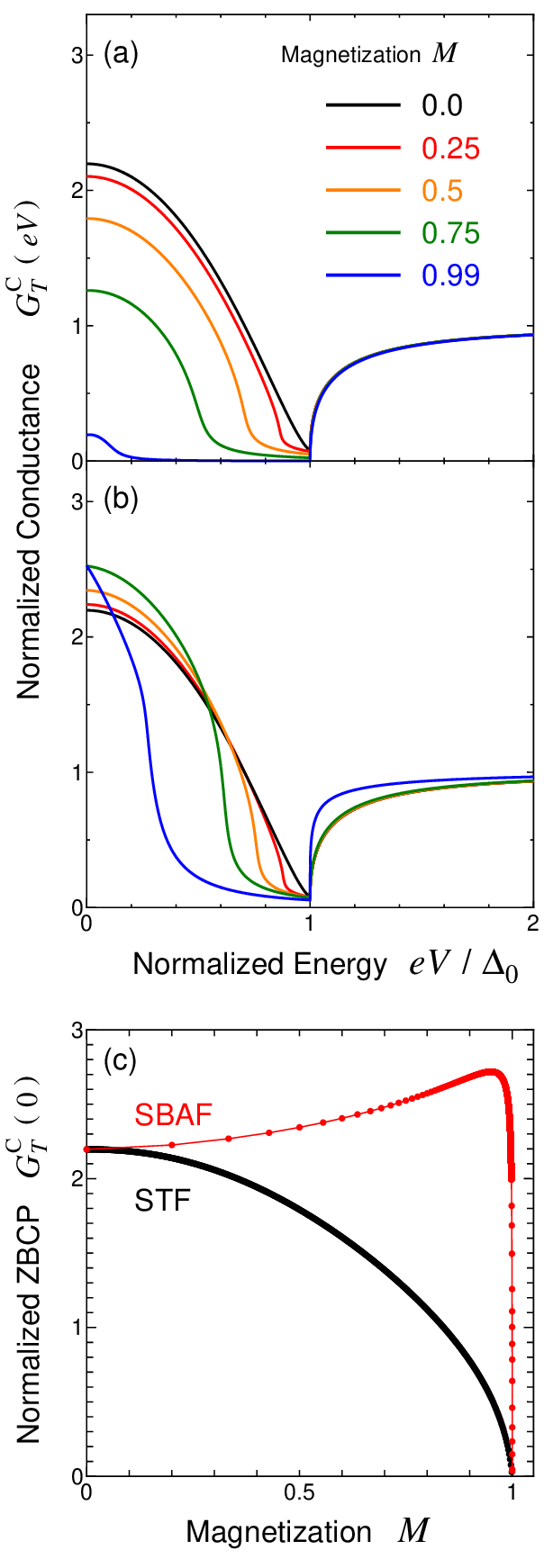}
\end{center}
\caption{ Normalized conductance spectra of the charge current $G_{T}^{C}(eV)$ in the tunneling limit, $Z=5$ 
for STF/I/TS junction (a) and for SBAF/I/TS junction (b). 
And in (c), height of ZBCPs is plotted as a function of the magnetization $M$ for STF (red line) and for SBAF (black line).}
\label{fig6}
\end{figure}

It has been shown theoretically that the ZBCP in F/I/S junction 
would be useful for measuring the magnetization of ferromagnet\cite{ev1,ev2}. 
In here, we study the validity of the ZBCP for the distinction of ferromagnets. 
Figure \ref{fig6} shows the conductance $G_{T}^{C}(eV)$ for the junction in the tunneling limit $Z=5$. 
The ZBCPs seen in both STF and SBAF cases are attributed to the anisotropy of the pair potential of 
$p$-wave superconductor. 
For STF case, the previous results\cite{ev1} have been reproduced(Fig.\ref{fig6}(a)). 
In contrast, there are some differences for SBAF case. 
Especially, it is found that the conductance near $eV=0$ increases slightly 
with increasing $M$ (Fig.\ref{fig6}(b)). 
This opposite behavior can be seen more clearly in the $M$ dependence of ZBCPiFig.\ref{fig6}(c). 
With increasing $M$, in contrast to the monotonically decreasing behavior 
of STF case, the ZBCP in SBAF case increases up to a certain value of $M$ in strong 
ferromagnetic regime and then, suddenly decreases toward the half metallic limit 
where the ZBCPs in both cases are suppressed perfectly. 
Cause of this opposite behavior could be reduced to the definition of normalization way 
since the magnitude of ZBCP being a constant value in non-normalized case 
depends on the conductance as the superconductor is in normal state. 
There is other definition of normalization by using the AR critical angle measured in the ferromagnet side\cite{ev1,ev2}. 
However, in that case, the AR critical angle itself depends on and is controlled by the magnitude of $M$, 
as a results, even the normalization depends on $M$. 
Accordingly, in order to avoid the influence of $M$, 
we alternatively calculate an angle averaged conductance 
defined as in the following,
\begin{figure}
\begin{center}
\includegraphics[width=8.5cm]{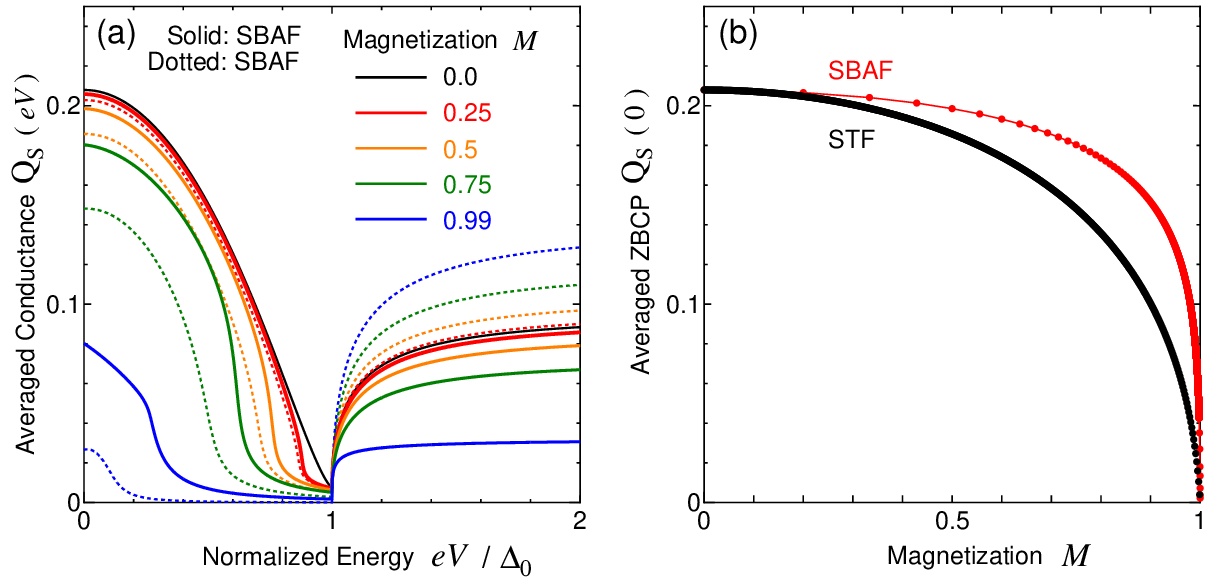}
\end{center}
\caption{ Angle averaged conductance spectra $Q_{S}$ as a function 
of magnetization (a) and ZBCPs vs. magnetization strength (b) in the tunneling limit $Z= 5$. 
Here, the conductance for SBAF case is indicated as solid line and for STF case is as dotted line. }
\label{fig7}
\end{figure}
\begin{figure}
\begin{center}
\includegraphics[width=8.5cm]{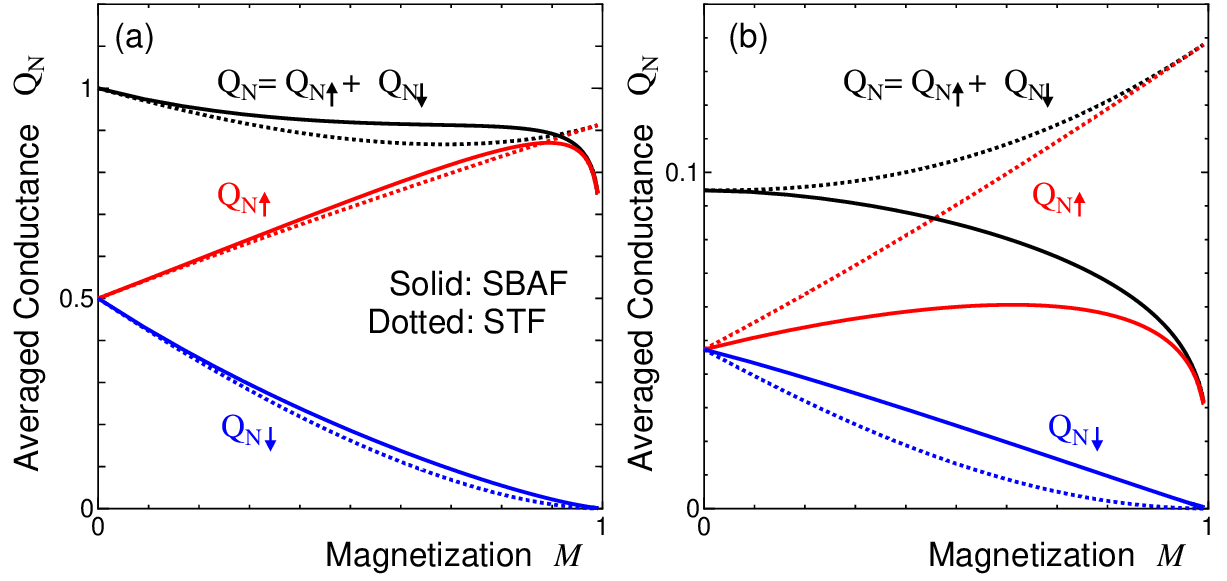}
\end{center}
\caption{ Magnetization dependence of angle averaged normal conductance $Q_{N}$, $Q_{N,\uparrow}$, 
and $Q_{N,\downarrow}$ for SBAF case (solid line) and STF case (dotted line) in the tunneling limit for the case of 
the superconductor is in normal state, (a) for $Z=0$, (b) for $Z=5$.}
\label{fig8}
\end{figure}
\begin{align}
\displaystyle
Q_{S}&=Q_{S,\uparrow}+Q_{S,\downarrow} \nonumber \\
Q_{S,\sigma}&=\frac{\int^{\pi/2}_{\pi/2} d\theta_{S} \cos\theta_{S} 
P_{\sigma}G_{S,\sigma}^{C}}{\int^{\pi/2}_{\pi/2} d\theta_{S} \cos\theta_{S}}.  
\notag
\end{align}

We show the calculated results of the angle averaged conductance $Q_{S}$ 
in Fig.\ref{fig7} which, in both STF and SBAF cases, show same tendency to decrease as increasing $M$(Fig.\ref{fig7}(a)). 
Similarly, the ZBCP is decreasing function of $M$(Fig.\ref{fig7}(b)). 
It is also shown that the reduction ratio differs in each of both cases 
as same as that in metallic limit. Thus, the opposite behavior seen in normalized conductance 
would reduce to the conductance in normal state. 
Therefore, it is noticed that the conductance of the junction for the superconductor being in normal state 
play an important role on our attention for two different ferromagnetisms. 

In order to clarify the difference between STF case and SBAF case more, we calculate the conductance in 
ferromagent/normal metal (F/I/N) junction 
for both in metallic and in tunneling limits. 
The angle averaged conductance in F/I/N junction $Q_{N}=\sum_{\sigma}Q_{N,\sigma}$ is defined in similar way to that in F/I/S junction 
replacing $G_{S,\sigma}^{C}$ by $G_{N,\sigma}^{C}$. 
The calculated results of $Q_{N,\sigma}$ for both $Z_{0}=0$ and $Z_{0}=5$ 
are shown in Fig.\ref{fig8}.
The angle resolved conductance $G_{N,\sigma}^{C}$ for $\sigma$-spin is rewritten by 
$G_{N,\sigma}^{C}=4\cos\theta_{S}\tilde{\lambda}_{\sigma}/
((\cos\theta_{S}+\tilde{\lambda}_{\sigma})^2 + Z_{0,\sigma}^2)$
where $\tilde{\lambda}_{\sigma}=\sqrt{\cos^{2}\theta_{S} + \rho\chi}$ in STF/I/N 
and $\tilde{\lambda}_{\sigma}
=\gamma^{-\rho/2}\sqrt{\cos^2\theta_{S} + (\gamma^{\rho/2}-1) }$ 
in SBAF/I/N junctions. 
Here, we mention properties of $M$-dependence of $G_{N,\sigma}^{C}$ through $\chi$ or $\gamma$ in advance of descriptions 
about $Q_{N}$. In STF/I/N junction, the $G_{N,\uparrow}^{C}$ increases following growth of the magnetization, i.e., 
with increasing $\chi$ since the gain of Fermi energy due to the band shift is larger than the Fermi surface 
effect\cite{ev1} acting as an effective barrier between STF and normal metal, 
under the conservation of the momentum along $y-$direction. 
On the other hand, because there is no Fermi energy gain from spread of the band width due to the effective mass 
mismatch in SBAF and the influence of the effective barrier arising from the Fermi surface effect becomes stronger with the increase of $\gamma$, 
the $G_{N,\uparrow}^{C}$ in SBAF/I/N junction decreases with increasing $\gamma$ and become zero in the limit of $\gamma\rightarrow\infty$. 
$G_{N,\downarrow}^{C}$ for both STF and SBAF cases decreases 
with increasing the magnetization caused by $\chi$ or $\gamma$. 

In the metallic limit $Z_{0}=0$ (Fig.\ref{fig8}(a)), 
it is found that the $Q_{N,\uparrow}$ in STF/I/N junction increases 
with increasing $M$ in contrast to $Q_{N,\downarrow}$ decreasing toward zero in half metal state. 
In this case, $M$ is given directly as $M=\chi$. 
Thus, the total conductance $Q_{N}= Q_{N,\uparrow}+ Q_{N,\downarrow}$ is reduced slightly by the Fermi surface effect 
with increasing $M$ up to $\sim 0.7$. 
In SBAF/I/N junction, we can see similar behavior in $Q_{N,\uparrow(\downarrow)}$. 
The increase of $Q_{N,\uparrow}$ is owing to  $P_{\uparrow}$ which is an increasing function of $\gamma$. 
However, near the half metallic limit, 
$Q_{N,\uparrow}$ reduces rapidly reflecting the behavior of $G_{N,\uparrow}$ which is a decreasing function of $M$ 
toward zero at $M=1(\gamma=\infty)$ as mentioned above. 
Thus, as shown in Fig.\ref{fig4}, the $G_{T}^{S}(eV)$ in SBAF/I/S junction decreases slowly with increasing $M$ with comparing to that in STF/I/S junction. 
The difference between STF and SBAF becomes more clearly in the tunneling limit $Z=5$(Fig.\ref{fig8}(b)). 
With increasing $M$, $Q_{N,\sigma}$ in STF case varies in rapidly rather than that in SBAF case. 
This is a difference of a barrier effect felt by particles with $\sigma$-spin in each cases. 
The barrier potential simply becomes relatively lower for particles with $\uparrow$-spin and higher for particles with 
$\downarrow$-spin in the STF case due to the rigid Fermi energy shift. However, the particles in SBAF directly feel 
the barrier potential because there is no shift of the Fermi energy. 
Thus, in SBAF/I/N junction, 
the increase of the magnitude of $Q_{N,\uparrow}$ due to $P_{\uparrow}$ is suppressed by the Fermi surface effect 
and barrier potential and then, $Q_{N,\uparrow}$ is 
getting lower with increasing $M$ in contrast to the STF case. 
Therefore, the $Q_{N}$ in SBAF/I/N junction shows the 
opposite behavior of that in STF/I/N junction. 
As a result, the normalized conductance $G_{T}^{S}(eV)$ 
in SBAF/I/S junction 
increases due to the reduction of the $Q_{N}$ depending on $M$ 
(Fig.\ref{fig6}(b)-(c)). Indeed, as shown in Fig.\ref{fig7}, 
the angle averaged conductance $Q_{S}(eV)$s for both STF and SBAF 
case show same trend on varying $M$. 
Thus, it can be conclude that the measurement 
of $Q_{N}$ will be also useful to identify the STF and SBAF. 
However, we emphasize that the measurement of ZBCP originated from 
ZABS is more powerful probe 
to investigate ferromagnet than that of $Q_{N}$.
Because, two $Q_{N}$s seemingly show drastically different behavior depending on $M$ 
for enough large $Z_{0}$ (Fig.\ref{fig8}(b)), 
by carefully looking of the figure, differences of each values of $Q_{N}$s are not so large for same $M$ 
except strong ferromagnetic regime. 
Therefore, it seems that an experimental distinction will become more difficult on measurement of $Q_{N}$. 
The ZBCP is getting more clear for larger $Z_{0}$, 
then which can be expected to play a role of good manifestation of the difference of STF and SBAF.

\section{\label{sec4}
Summary}

In summary, we have derived a formula of the tunneling conductance 
in ferromagnet/ferromagnetic-insulator/superconductor 
with antiparallel spin pairing junction by extending our previous theory 
for standard Stoner ferromagnet (STF) so as to include spin-band asymmetry ferromagnet (SBAF) 
originated from effective mass mismatch between particles with opposite spins. 
Applying the formulation to 
ferromaget/insulator/$p$-wave superconductor junctions, 
differences between pure STF and pure SBAF have been investigated intensively. 
We found that, with growing the magnetization, 
the difference becomes clear in tunneling conductance. 
The clarity of difference between STF and SBAF depends on the way 
of normalization of conductance and comes out more clearly 
in ZBCP near half-metallic limit. 
The obtained results suggest that the measurement of ZBCP may be useful 
for discriminating mechanism of ferromagnetism. 

Although our formulation includes the ferromagnetic insulator, 
we have studied only the normal insulating barrier case in this paper. 
The spin-filtering effect have been expected 
in the ferromagnetic insulator\cite{ev1} or 
in ferromagnet given by the effective mass mismatch\cite{gae3}.
Then, as an interesting future problem  we will study extensively  
the spin-filtering effect in junctions of including both ferromagnetic 
insulator and mass mismatch ferromagnet connected to superconductors of $s$-, 
$d$-wave and broken time reversal symmetry pairing states. 
Moreover, it will be an important issue that the proximity effect is taken into account to 
the present formulation by carrying out the self-consistent 
calculation of the pairing potential in order to analyze the actual experiments. 
Indeed, the ZBCP have been observed in tunneling experiment of 
F/I/$d$-wave superconductor junction\cite{sawa}. And also, ZBCP in Sr$_{2}$RuO$_{4}$ 
junction has been observed\cite{sk3}, then, tunneling spectroscopy of F/I/Sr$_{2}$RuO$_{4}$ junction 
seems to be realized in near future. Our conductance formula can apply to such situations 
easily and get comparable results to experimental one.

\end{document}